\documentclass[useAMS]{mn2e}
 \usepackage{times}
 \usepackage{graphicx}
 \usepackage{amssymb}

 \title[Flipping minor bodies: comet 96P/Machholz 1]
       {Flipping minor bodies: what comet 96P/Machholz 1 can tell us about the orbital evolution of extreme trans-Neptunian objects 
                               and the production of near-Earth objects on retrograde orbits}

 \author[C. de la Fuente Marcos, R. de la Fuente Marcos and S. J. Aarseth]
        {Carlos~de~la~Fuente~Marcos,$^{1}$\thanks{E-mail: nbplanet@fis.ucm.es}
         Ra\'ul~de~la~Fuente Marcos$^{1}$
         and
         Sverre J. Aarseth$^2$ \\
         $^1$Universidad Complutense de Madrid,
             Ciudad Universitaria, E-28040 Madrid, Spain \\
         $^2$Institute of Astronomy, University of Cambridge, 
             Madingley Road, Cambridge CB3 0HA, UK}
 \date{Accepted 2014 October 22.
       Received 2014 October 16;
       in original form 2014 September 23}
 \pubyear{2015}
 
\begin{document}
  \maketitle

  \begin{abstract}
     Nearly all known extreme trans-Neptunian objects (ETNOs) have 
     argument of perihelion close to 0\degr. An existing observational 
     bias strongly favours the detection of ETNOs with arguments of
     perihelion close to 0\degr and 180\degr yet no objects have been 
     found at 180\degr. No plausible explanation has been offered so far 
     to account for this unusual pattern. Here, we study the dynamical 
     evolution of comet 96P/Machholz 1, a bizarre near-Earth object 
     (NEO) that may provide the key to explain the puzzling clustering 
     of orbits around argument of perihelion close to 0\degr recently 
     found for the population of ETNOs. Comet 96P/Machholz 1 is 
     currently locked in a Kozai resonance with Jupiter such that the 
     value of its argument of perihelion is always close to 0\degr at 
     its shortest possible perihelion (highest eccentricity and lowest 
     inclination) and about 180\degr near its shortest aphelion 
     (longest perihelion distance, lowest eccentricity and highest 
     inclination). If this object is a dynamical analogue (albeit 
     limited) of the known ETNOs, this implies that massive perturbers 
     must keep them confined in orbital parameter space. Besides, its 
     future dynamical evolution displays orbital flips when its 
     eccentricity is excited to a high value and its orbit turns over 
     by nearly 180\degr, rolling over its major axis. This unusual 
     behaviour, that is preserved when post-Newtonian terms are 
     included in the numerical integrations, may also help understand 
     the production of NEOs on retrograde orbits. 
  \end{abstract}


  \section{Introduction}
     Two Solar system discoveries have recently puzzled the astronomical community: the existence of near-Earth objects (NEOs, perihelion 
     distance less than 1.3 au and, if a comet, orbital period less than 200 yr) moving on retrograde orbits and the clustering of orbits 
     around argument of perihelion, $\omega$, close to 0\degr for extreme trans-Neptunian objects (ETNOs, semimajor axis greater than 150 au 
     and perihelion distance greater than 30 au).

     There are only 11 known retrograde NEOs: three asteroids -- (343158) 2009 HC$_{82}$, 2007 VA$_{85}$ and 2014 PP$_{69}$ -- and eight
     comets (55P/Tempel-Tuttle, 1P/Halley, P/2005 T4 (SWAN), C/2010 L5 (WISE), 273P/Pons-Gambart, C/2001 W2 (BATTERS), 109P/Swift-Tuttle and 
     161P/Hartley-IRAS). The number of known near-Earth asteroids (NEAs) is 11\,490 and the number of near-Earth comets (NECs) is 165.
     Retrograde objects represent a tiny fraction, 0.094 per cent, of the NEO population. Numerical integrations carried out by Greenstreet
     et al. (2012) unveiled a dynamical mechanism capable of inducing ordinary asteroid orbits to flip to a retrograde configuration while 
     trapped in the 3:1 mean motion resonance with Jupiter near 2.5 au. These authors predict that nearly 0.1 per cent of the NEO 
     population could follow retrograde orbits.
     
     Nearly all known ETNOs have $\omega$ close to 0\degr (the average value is -31\degr$\pm$50\degr, see a discussion in Trujillo \& 
     Sheppard 2014; de la Fuente Marcos \& de la Fuente Marcos 2014b). An existing observational bias strongly favours the detection of 
     ETNOs at $\omega$ close to 0\degr and 180\degr yet no objects have been found at 180\degr (Trujillo \& Sheppard 2014; de la Fuente 
     Marcos \& de la Fuente Marcos 2014b). No plausible explanation has been offered so far to account for the apparent lack of objects with 
     $\omega$ around 180\degr. The perplexing puzzle posed by the apparent lack of such objects among known members of the ETNO population 
     is not the only unusual pattern displayed by the orbital parameters of this interesting group of trans-Plutonian objects (see fig. 3 in 
     de la Fuente Marcos \& de la Fuente Marcos 2014b). The orbital elements of these objects (see Table \ref{ETNOs}) exhibit conspicuous 
     clustering around $e\sim$0.80 (0.82$\pm$0.06) and $i\sim$20\degr (18\degr$\pm$7\degr). The very high mean value of the eccentricity 
     could be attributed, in principle, to an observational bias: for this population, objects with the shortest perihelia (and, therefore, 
     the highest eccentricity) should be the easiest to find. The moderately high average inclination is however surprising: objects with 
     inclinations close to 0\degr are nearly twice more likely to be found than those at 20\degr yet none has been discovered (2010~VZ$_{98}$ 
     has $i$=4\fdg51, see Table \ref{ETNOs}). This orbital behaviour does not fit easily within the well documented Kozai dynamics beyond 
     Neptune (Gallardo, Hugo \& Pais 2012). The orbital distributions of the known ETNOs are unlike those of objects part of the 
     trans-Neptunian belt as characterised by e.g. Fern\'andez (1980). The unexpected clustering in $\omega$ around 0\degr but not around 
     180\degr has been tentatively associated with the Kozai resonance (Kozai 1962) by Trujillo \& Sheppard (2014) and de la Fuente Marcos \& 
     de la Fuente Marcos (2014b). 
%
%-------------------------------------------------------------------------------------------------------------------------- Orbital elements
%
      \begin{table*}
        \centering
        \fontsize{8}{11pt}\selectfont
        \tabcolsep 0.25truecm
        \caption{Various orbital parameters ($\varpi = \Omega + \omega$, $\lambda = \varpi + M$) for the 13 known ETNOs
                 (Epoch: JD 245\,7000.5 that corresponds to 0:00 \textsc{ut} on 2014 December 9. J2000.0 ecliptic and equinox.
                  Source: JPL Small-Body Database.)
                }
        \begin{tabular}{lcccccccc}
          \hline
             Object             & $a$ (au)    & $e$        & $i$ (\degr) & $\Omega$ (\degr) & $\omega$ (\degr) & $\varpi$ (\degr) & $\lambda$ (\degr)
                                & $Q$ (au)    \\
          \hline
     (82158) 2001 FP$_{185}$    & 222.8895152 & 0.84638769 & 30.76961    & 179.31663        &    6.83700       & 186.15400        & 187.34300
                                &  411.540    \\
             (90377) Sedna      & 524.3945961 & 0.85489532 & 11.92862    & 144.54452        &  -48.71210       &  95.83240        &  93.99530
                                &  972.697    \\
    (148209) 2000~CR$_{105}$    & 230.1151596 & 0.80777192 & 22.70702    & 128.23435        &  -42.84210       &  90.48600        &  90.48600
                                &  415.996    \\
             2002~GB$_{32}$     & 211.8628339 & 0.83318809 & 14.17959    & 176.99786        &   36.89740       & 213.98600        & 213.98600
                                &  388.384    \\
             2003~HB$_{57}$     & 162.3925720 & 0.76545417 & 15.48811    & 197.84847        &   10.66603       & 209.54300        & 209.54300
                                &  286.697    \\
             2003~SS$_{422}$    & 195.9581432 & 0.79878068 & 16.80711    & 151.10976        & -149.98800       &   1.84113        &   1.84113
                                &  352.486    \\
             2004~VN$_{112}$    & 328.8226182 & 0.85605608 & 25.53759    &  66.03781        &  -32.77670       &  33.57160        &  33.57160
                                &  610.313    \\
             2005~RH$_{52}$     & 151.9398031 & 0.74329346 & 20.46837    & 306.19632        &   32.51841       & 340.91900        & 340.91900
                                &  264.876    \\
             2007~TG$_{422}$    & 518.1738582 & 0.93132860 & 18.58061    & 112.97697        &  -74.16600       &  39.08750        &  39.08750
                                & 1000.760    \\
             2007~VJ$_{305}$    & 190.7687814 & 0.81549271 & 11.99449    &  24.38349        &  -21.51450       &   4.08605        &   4.08605
                                &  346.339    \\
             2010~GB$_{174}$    & 370.1872857 & 0.86864073 & 21.53245    & 130.58694        &  -12.34630       & 118.24100        & 121.41800
                                &  691.747    \\
             2010~VZ$_{98}$     & 155.2582279 & 0.77896305 &  4.50914    & 117.46709        &  -46.15010       &  71.31700        &  68.90210
                                &  276.199    \\
             2012~VP$_{113}$    & 263.1193815 & 0.69400129 & 24.02419    &  90.86971        &  -66.16680       &  24.70290        &  27.72750
                                &  445.725    \\
          \hline
        \end{tabular}
        \label{ETNOs}
      \end{table*}
%
%-------------------------------------------------------------------------------------------------------------------------------------------
%

     Here, we show that the orbital evolution of the bizarre NEC 96P/Machholz 1 may explain naturally the clustering of orbits around 
     argument of perihelion close to 0\degr observed for ETNOs and also provide another viable dynamical evolutionary pathway to produce 
     retrograde NEOs. This paper is organized as follows. In Section 2, we briefly outline our numerical model. In Section 3, we review 
     what is currently known about this object. The classical dynamical evolution of comet 96P/Machholz 1 is presented in Section 4. Orbital 
     flips, including relativistic issues, are analysed in Section 5. The relevance of our findings within the context of the ETNO 
     population is discussed in Section 6. Section 7 summarizes our conclusions.
%
%----------------------------------------------------------------------------------------------------------------------------------- TABLE I
%----------------------------------------------------------------------------------------------------- Orbital elements comet 96P/Machholz 1
%
     \begin{table}
      \fontsize{8}{11pt}\selectfont
      \tabcolsep 0.15truecm
      \caption{Heliocentric Keplerian orbital elements of comet 96P/Machholz 1. Values include the 1$\sigma$ uncertainty. The orbit is 
               computed at Epoch JD 245\,6541.5 that corresponds to 0:00 \textsc{ut} on 2013 September 6 (J2000.0 ecliptic and equinox. 
               Source: JPL Small-Body Database.)
              }
      \begin{tabular}{ccc}
       \hline
        Semimajor axis, $a$ (au)                               & = &   3.033 93972$\pm$0.000 00002 \\
        Eccentricity, $e$                                      & = &   0.959 21183$\pm$0.000 00005 \\
        Inclination, $i$ ($^{\circ}$)                          & = &  58.312 21$\pm$0.000 04       \\
        Longitude of the ascending node, $\Omega$ ($^{\circ}$) & = &  94.323 236$\pm$0.000 012     \\
        Argument of perihelion, $\omega$ ($^{\circ}$)          & = &  14.757 75$\pm$0.000 02       \\
        Mean anomaly, $M$ ($^{\circ}$)                         & = &  77.992 761$\pm$0.000 003     \\
        Perihelion, $q$ (au)                                   & = &   0.123 74885$\pm$0.000 00014 \\
        Aphelion, $Q$ (au)                                     & = &   5.944 13060$\pm$0.000 00005 \\
        Comet total magnitude, $H$ (mag)                       & = &  13.3                         \\
       \hline
      \end{tabular}
      \label{elements}
     \end{table}
%
%-------------------------------------------------------------------------------------------------------------------------------------------
%
  \section{Numerical model}
     The numerical integrations of the orbits of comet 96P/Machholz 1 studied here were performed with the Hermite integrator (Makino 1991;
     Aarseth 2003), in a model Solar system which takes into account the perturbations by eight major planets and treats the Earth--Moon 
     system as two separate objects; it also includes the barycentre of the dwarf planet Pluto--Charon system and the 10 most massive 
     asteroids of the main belt, namely (1) Ceres, (2) Pallas, (4) Vesta, (10) Hygiea, (31) Euphrosyne, 704 Interamnia (1910 KU), 511 Davida 
     (1903 LU), 532 Herculina (1904 NY), (15) Eunomia and (3) Juno (for further details, see de la Fuente Marcos \& de la Fuente Marcos 
     2012). Results in the figures have been obtained using initial conditions (positions and velocities in the barycentre of the Solar 
     system) provided by the Jet Propulsion Laboratory (JPL) \textsc{horizons} system (Giorgini et al. 1996; Standish 1998) and referred to 
     the JD 245\,7000.5 epoch which is the $t$ = 0 instant. In addition to the calculations completed using the nominal orbital elements in 
     Table \ref{elements}, we have performed 50 control simulations with sets of orbital elements obtained from the nominal ones within the 
     accepted uncertainties (3$\sigma$).
%
%---------------------------------------------------------------------------------------------------------------------------------- FIGURE 1
%
     \begin{figure}
       \centering
        \includegraphics[width=\linewidth]{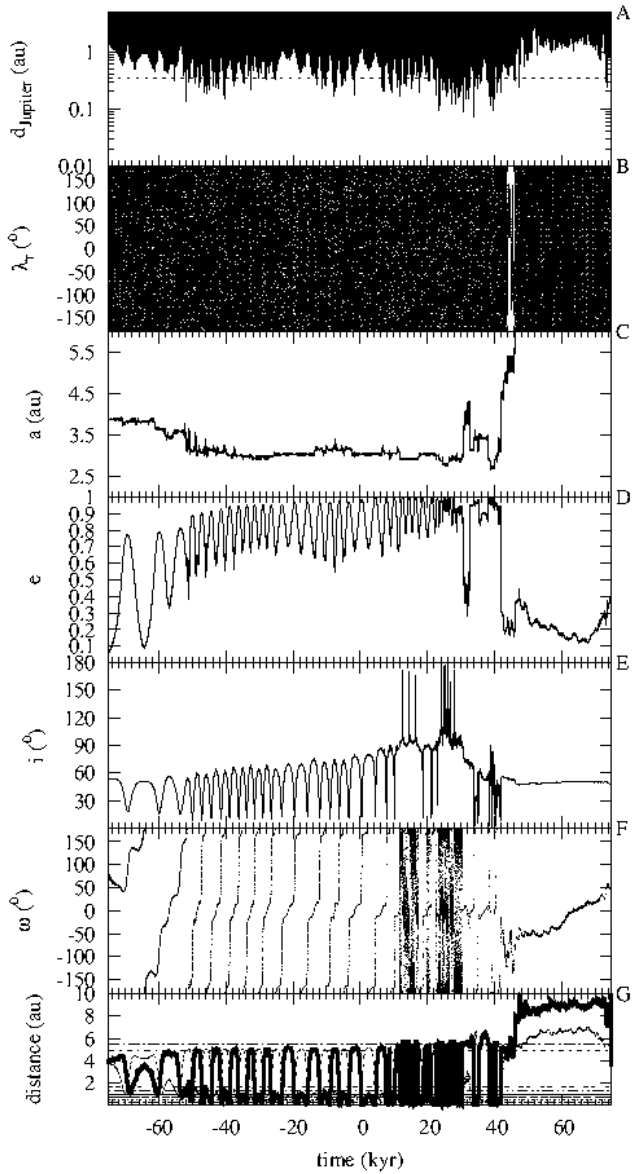}
        \caption{Time evolution of various parameters for the nominal orbital solution of comet 96P/Machholz 1. The distance from Jupiter 
                 (panel A); the value of the Hill sphere radius of Jupiter, 0.35 au, is displayed. The resonant angle, $\lambda_{\rm r} = 
                 \lambda - \lambda_{\rm Jupiter}$ (panel B) for the orbit in Table \ref{elements}. The orbital elements $a$ (panel C), $e$ 
                 (panel D), $i$ (panel E) and $\omega$ (panel F). The distance to the descending (thick line) and ascending nodes (dotted 
                 line) is in panel G. Planetary perihelion and aphelion distances are also shown.
                }
        \label{comet}
     \end{figure}
%
%-------------------------------------------------------------------------------------------------------------------------------------------
%

  \section{Comet 96P/Machholz 1, bizarre from every angle}
     Comet 96P/Machholz 1 was discovered on 1986 May 12 by D. E. Machholz observing with 29$\times$130 binoculars from Loma Prieta, 
     California (Machholz, Morris \& Hale 1986) and confirmed the following days by S. Morris and A. Hale observing from near Mt. Wilson, 
     California. It was soon clear that its orbit was very unusual, travelling closer to the Sun than any known planet, (at that time) 
     asteroid or comet (Green et al. 1990), and displaying higher than expected activity at aphelion (Sekanina 1990). With an orbital period 
     of 5.28 yr and a perihelion, $q$, of just 0.12 au, its eccentricity, $e$, is 0.96, and its inclination, $i$, is significant at 58\fdg31. 
     Its aphelion, $Q$, is beyond Jupiter at 5.94 au. Its current orbit is based on 1090 observations with a data-arc span of 9642 d.
     Babadzhanov \& Obrukov (1992), Gonczi, Rickman \& Froeschl\'e (1992), Jones \& Jones (1993), Ka\v{n}uchov\'a \& Neslu\v{s}an (2007) and
     Neslu\v{s}an, Ka\v{n}uchov\'a \& Tomko (2013) suggested that this object is the parent body of the Quadrantid meteors and perhaps other
     meteor showers. Jenniskens et al. (1997) and Williams \& Collander-Brown (1998) considered that the Quadrantids do not originate from 
     comet 96P/Machholz 1 although it could be actively releasing meteoroids (Ohtsuka, Nakano \& Yoshikawa 2003). In fact, it could be the 
     ancestor of the Marsden and Kracht groups of sunskirting comets (Sekanina \& Chodas 2005). Its spectrum is very unusual (Langland-Shula 
     \& Smith 2007; Schleicher 2008). Scenarios for the origin of this atypical object include being an interstellar comet, formation in the 
     ourskirts of the Oort Cloud, or having a standard origin but being chemically altered due to its frequent excursions inside Mercury's 
     orbit (see e.g. Schleicher 2008). Levison \& Dones (2014) have shown that 96P/Machholz 1 is trapped in a Kozai resonance with Jupiter 
     in which its orbital eccentricity oscillates between 0.6 and almost 1.0 and its inclination varies between roughly 10\degr and 80\degr 
     (see their fig. 31.2). The value of $\omega$ for this comet is 14\fdg76 and when its eccentricity is close to 0.85, the inclination is 
     nearly 20\degr (see fig. 31.2 in Levison \& Dones 2014). These authors, as Gonczi et al. (1992) and others did, predict that this comet 
     will hit the Sun in less than 12 kyr.
 
  \section{Comet 96P/Machholz 1: orbital evolution}
\begin{figure*}
       \centering
        \includegraphics[width=\linewidth]{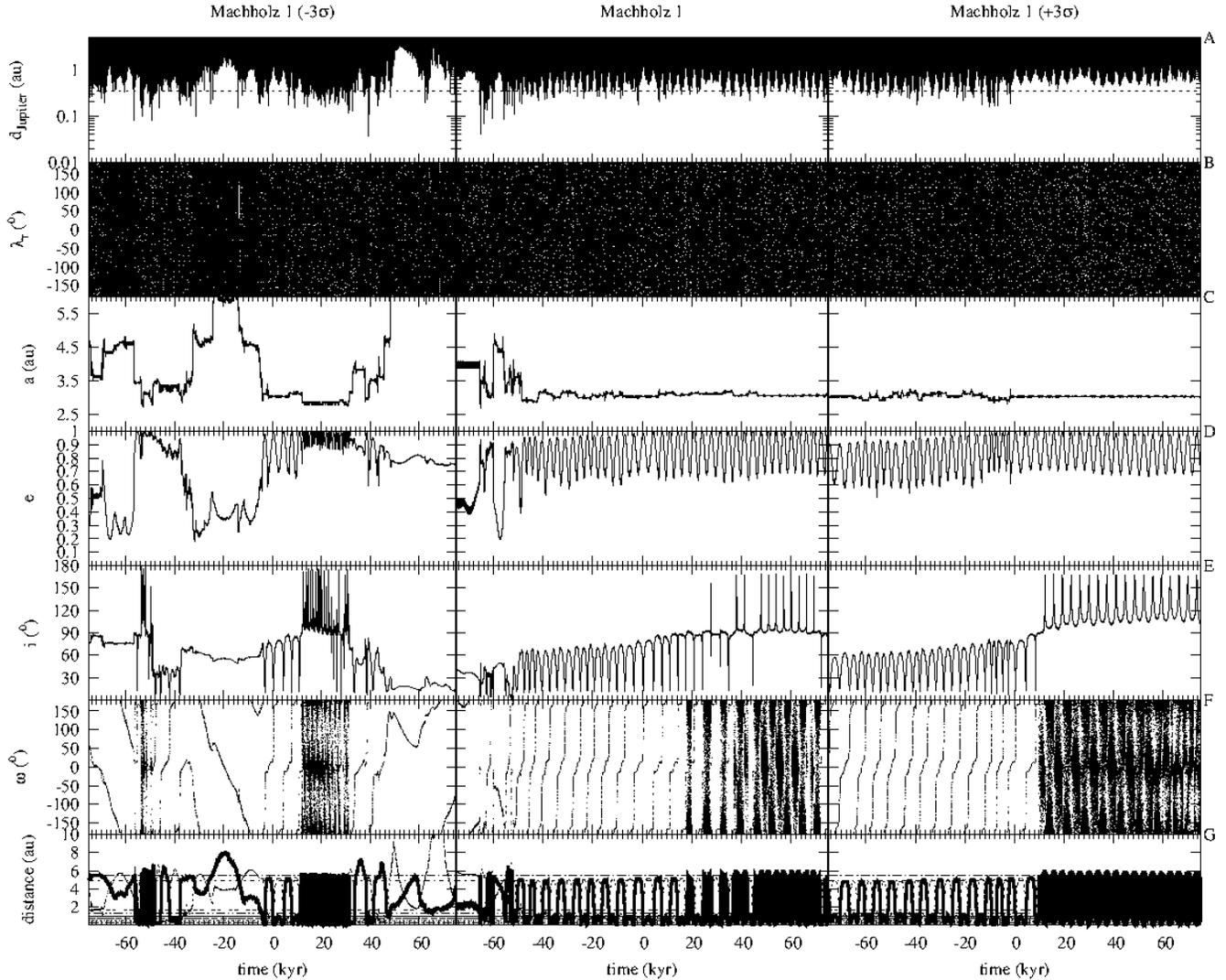}
        \caption{Comparative short-term dynamical evolution of various parameters for the nominal orbit of comet 96P/Machholz 1 as presented 
                 in Table \ref{elements} (central panels) and two representative examples of orbits that are most different from the nominal
                 one (see the text for details). The distance from Jupiter (A-panels); the value of the Hill sphere radius of Jupiter,
                 0.35 au, is displayed. The resonant angle, $\lambda_{\rm r}$ (B-panels). The orbital elements $a$ (C-panels), $e$ 
                 (D-panels), $i$ (E-panels) and $\omega$ (F-panels). The distances to the descending (thick line) and ascending nodes 
                 (dotted line) appear in the G-panels. Planetary aphelion and perihelion distances are also shown. The orbit labelled as 
                 `nominal' is arbitrarily close to the nominal one but not that of Fig. \ref{comet}.
                }
        \label{control96p}
     \end{figure*}

     Comet 96P/Machholz 1 is a well studied Jupiter-family comet. With a semimajor axis of 3.03 au, it is submitted to the 9:4 mean motion 
     resonance with Jupiter (see i.e. Ohtsuka et al. 2003). The time evolution of the osculating orbital elements of its nominal orbit is 
     displayed in Fig. \ref{comet}. We confirm that the object is currently trapped in a Kozai resonance with Jupiter. Because of this, the 
     eccentricity and inclination oscillate with the same frequency but out of phase (see panels D and E), when the value of the 
     eccentricity reaches its maximum the value of the inclination is the lowest and vice versa ($\sqrt{1 - e^2} \cos i =$ constant). In 
     a Kozai resonance, the apse and the node are in resonance with one another (Kozai 1962). The values of eccentricity and inclination are 
     coupled, and the value of the semimajor axis remains nearly constant (see panels C, D and E in Fig. \ref{comet}). The orbit of this 
     object is particularly chaotic and it is difficult to make reliable predictions beyond a few thousand years. All the integrated orbits
     give consistent results within the time interval (-2, 6.5) kyr. Its current Kozai resonant dynamical status is firmly established. As 
     an example, Fig. \ref{control96p} displays the short-term dynamical evolution of an orbit arbitrarily close to the nominal one (central 
     panels) and those of two illustrative worst orbits which are most different from the nominal one. The orbit labelled as `-3$\sigma$' 
     (left-hand panels) has been obtained by subtracting thrice the uncertainty from the orbital parameters (the six elements) in Table 
     \ref{elements}. It has the lowest values of $a$, $e$ and $i$ at the 3$\sigma$ level. In contrast, the orbit labelled as `+3$\sigma$' 
     (right-hand panels) was computed by adding three times the value of the uncertainty to the orbital elements in Table \ref{elements}. 
     This trajectory has the largest values of $a$, $e$ and $i$ (within 3$\sigma$). Close encounters with Jupiter within the Hill radius
     (see panel A) are very frequent with one of the nodes usually close to Jupiter (see panel G). With this orbital layout, the object is
     necessarily transient. We have neglected the role of non-gravitational forces in our simulations because the objective of this research
     is not the dynamics of the comet itself but its implications for the orbital evolution of other asteroidal bodies, NEOs and ETNOs. None
     of the control orbits computed here will drive the comet into the Sun as predicted by Levison \& Dones (2014) and others. The inclusion 
     in the calculations of (10) Hygiea, (31) Euphrosyne, 704 Interamnia (1910 KU) and 511 Davida (1903 LU) has a major impact on the 
     simulated evolution as the object becomes a transient, for about 1 kyr, co-orbital to some of them. Relatively close encounters with 
     all the previously mentioned asteroids are possible.
%
%-------------------------------------------------------------------------------------------------------------------------------------------
%
%
%-------------------------------------------------------------------------------------------------------------------------------------------
%

  \section{Flipping the orbit}
     Perhaps, the most striking feature in Figs \ref{comet} and \ref{control96p} (E-panels) is the dramatic flip from prograde to retrograde 
     and back again observed after integrating the orbit at least 12 kyr into the future. The actual instant of the first orbital flip 
     depends on the initial conditions but it is found in all the integrations. Some retrograde episodes are short (see Fig. \ref{comet}) 
     but longer events are possible (see Fig. \ref{control96p}). These surprising events can be understood within the context of the 
     eccentric Kozai mechanism (Lithwick \& Naoz 2011). In an attempt to explain the existence of counter-orbiting exoplanetary systems, Li 
     et al. (2014) have demonstrated that for an initially coplanar hierarchical three-body system with both eccentric inner and outer 
     orbits, the eccentricity of the inner orbit can be excited to high values and the orbit can flip by $\sim$180\degr, rolling over its 
     major axis. The sudden orbit flip is the result of the comet continuously losing angular momentum that is in return gained by the outer 
     perturber, Jupiter in this case. When the comet's angular momentum is very small, a slight gravitational kick during a close encounter 
     is enough to switch the direction of its orbit. 

\begin{figure}
       \centering
        \includegraphics[width=\linewidth]{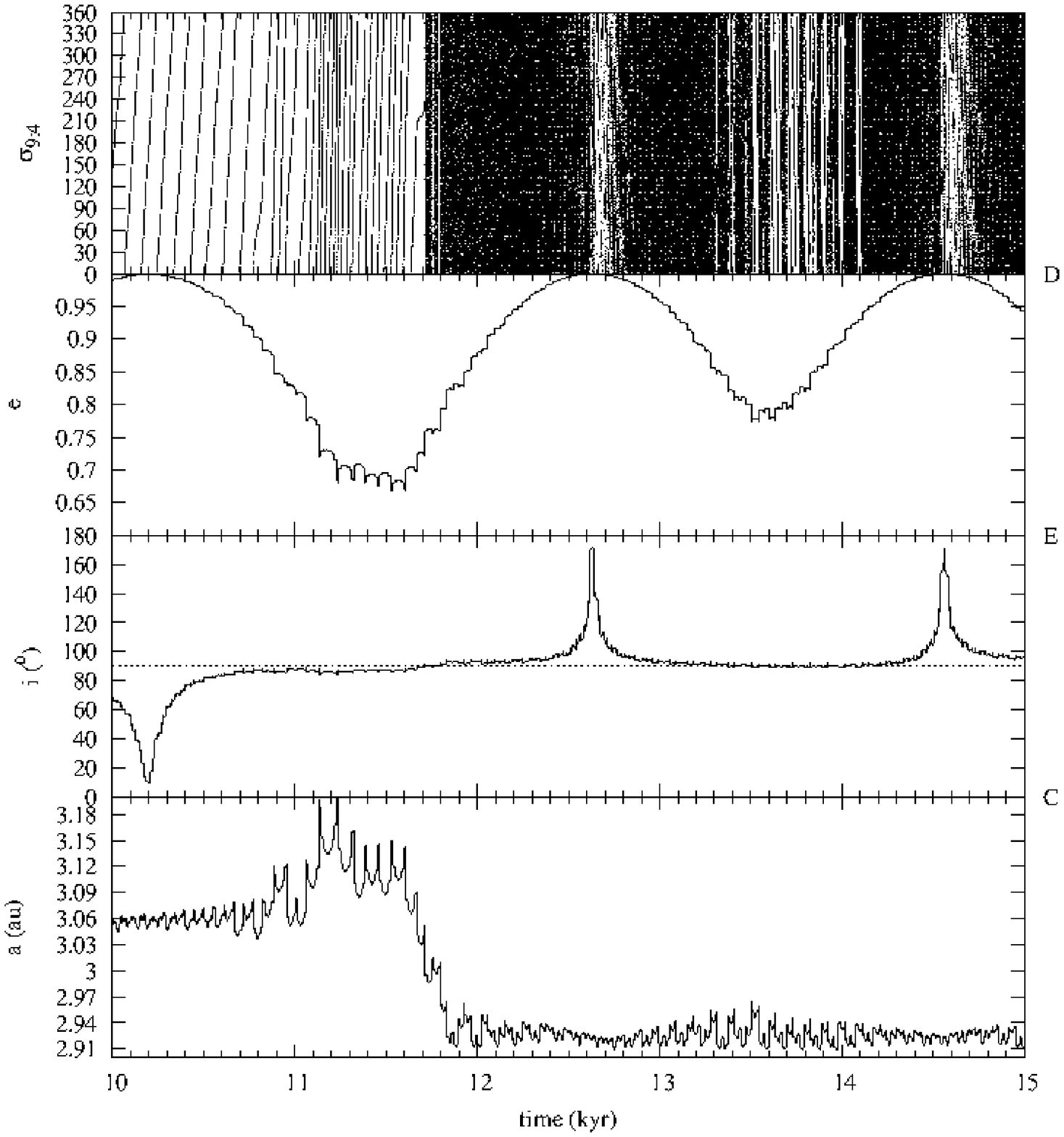}
        \caption{Resonant argument, $\sigma_{9:4}$, panels D, E and C (of Fig. \ref{comet}) restricted to the time interval (10, 15) kyr. 
                 The resonant argument associated to the 9:4 mean motion resonance with Jupiter becomes constant prior to the orbital flip 
                 from prograde to retrograde. The resonant argument is $\sigma_{\rm 9:4} = 9 \lambda_{\rm J} - 4 \lambda - 5 \varpi$, where
                 $\lambda_{\rm J}$ is the mean longitude of Jupiter, $\lambda$ is the mean longitude of comet 96P/Machholz 1, and
                 $\varpi = \Omega + \omega$ is its longitude of perihelion.
                }
        \label{94}
     \end{figure}

\begin{figure}
       \centering
        \includegraphics[width=\linewidth]{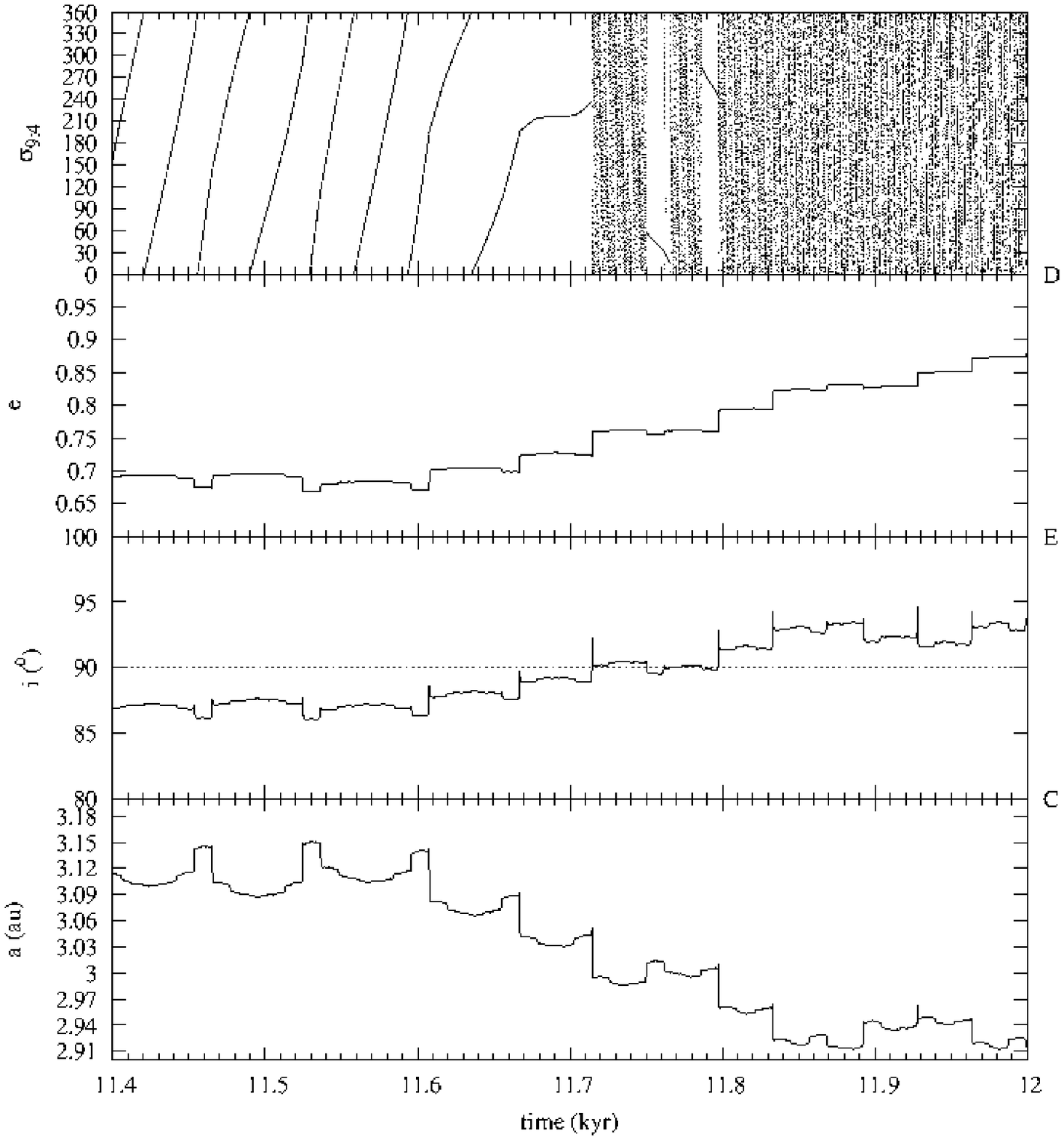}
        \caption{Same as Fig. \ref{94} but for the time interval (11.4, 12.0) kyr. 
                }
        \label{94z}
     \end{figure}

     The mechanism proposed by Greenstreet et al. (2012) to produce retrograde NEOs is different from the one presented here. In their
     simulations, the role of the 3:1 mean motion resonance (near $a$ = 2.5 au) is central to make the orbits retrograde. In our case, the 
     9:4 mean motion resonance with Jupiter is the one at work when the flip takes place at aphelion (see Figs \ref{94} and \ref{94z}). 
     Crossing the resonance triggers the orbital flip, see Fig. \ref{94z}. It is, therefore, another evolutionary pathway to produce 
     retrograde NEOs. The existence of multiple dynamical tracks running into the retrograde orbital domain can help to explain the 
     detection of high-velocity, rocky meteoroids on retrograde orbits (Borovi\v{c}ka et al. 2005). 
%
%---------------------------------------------------------------------------------------------------------------------------------- FIGURE 1
%
%
%-------------------------------------------------------------------------------------------------------------------------------------------
%
%
%---------------------------------------------------------------------------------------------------------------------------------- FIGURE 1
%
%
%-------------------------------------------------------------------------------------------------------------------------------------------
%

\begin{figure}
       \centering
        \includegraphics[width=\linewidth]{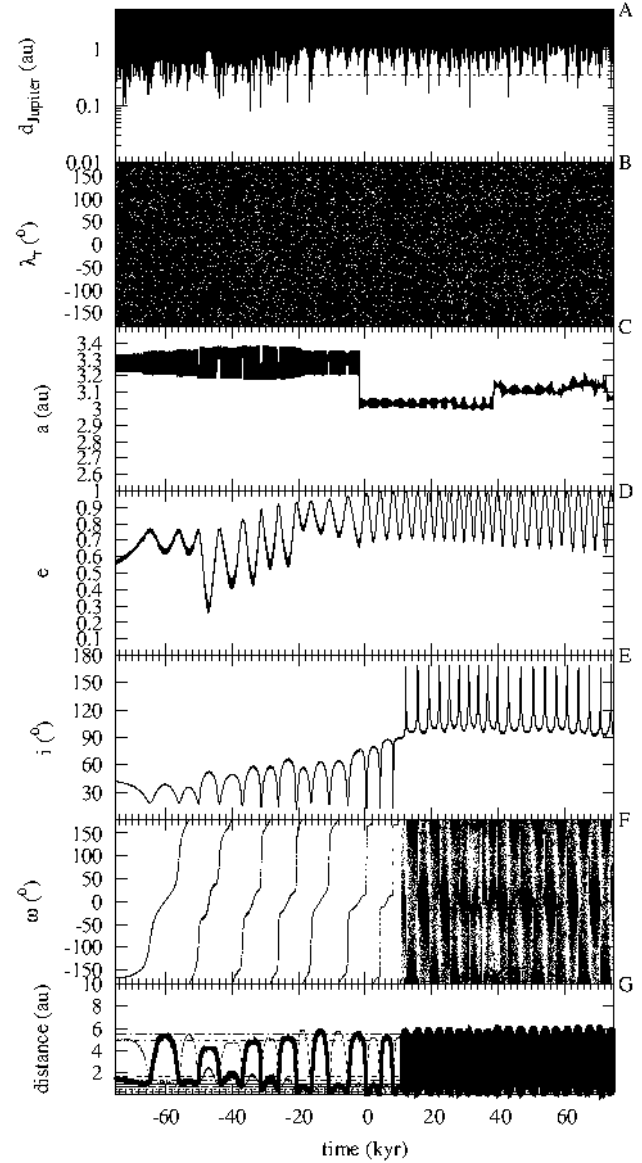}
        \caption{Same as Fig. \ref{comet} but for relativistic calculations. 
                }
        \label{cometr}
     \end{figure}

\begin{figure*}
       \centering
        \includegraphics[width=\linewidth]{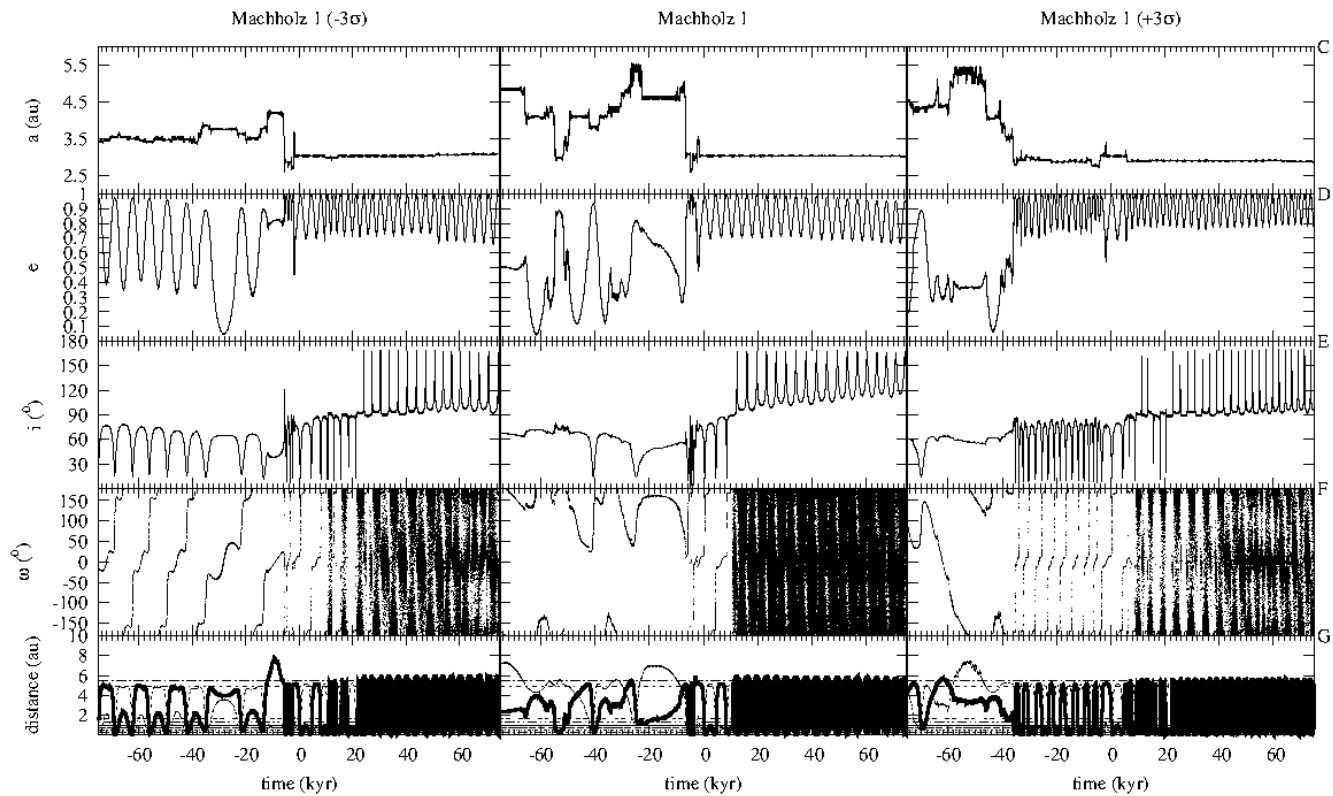}
        \caption{Comparative short-term dynamical evolution of various parameters for the nominal orbit of comet 96P/Machholz 1 as presented 
                 in Table \ref{elements} (central panels) and two representative examples of orbits that are most different from the nominal
                 one (see the text for details) in the post-Newtonian regime. The orbital elements $a$ (C-panels), $e$ (D-panels), $i$ 
                 (E-panels) and $\omega$ (F-panels). The distances to the descending (thick line) and ascending nodes (dotted line) appear 
                 in the G-panels. Planetary aphelion and perihelion distances are also shown. The orbit labelled as `nominal' is arbitrarily 
                 close to the nominal one but not that of Fig. \ref{cometr}. The orbits are as in Fig. \ref{control96p}.
                }
        \label{control96pr}
     \end{figure*}

     Relativistic effects, resulting from the theory of general relativity are not negligible when studying the long-term dynamical 
     evolution of minor bodies in the innermost part of the Solar system. Comet 96P/Machholz 1 has a low perihelion distance ($q$ = 0.12 
     au), comparable to those of the so-called relativistic asteroids (see table 2 in Benitez \& Gallardo 2008). For these objects, the 
     main effect is in the evolution of the argument of perihelion which has a direct impact on the Kozai mechanism. The inclusion of these 
     relativistic effects in $N$-body simulations is customarily approached within the framework of the post-Newtonian approximation (see 
     e.g. Aarseth 2007 and references therein) based on first-order expansion. Naoz et al. (2013) have found that the inclusion of 
     post-Newtonian contributions in the study of the Kozai mechanism in hierarchical three-body systems can, in some cases, suppress 
     eccentricity-inclination oscillations as the ones observed above when only Newtonian terms are considered in the integrations. Figure 
     \ref{cometr} shows that this is not the case here, the unusual behaviour found in the Newtonian case is preserved under the 
     post-Newtonian approximation. Further modifying the orbit as described in the previous section does not affect the overall results,
     see Fig. \ref{control96pr}. The duration of the retrograde episodes is, in fact, longer in the post-Newtonian case.
%
%---------------------------------------------------------------------------------------------------------------------------------- FIGURE 1
%
%
%-------------------------------------------------------------------------------------------------------------------------------------------
%
%
%-------------------------------------------------------------------------------------------------------------------------------------------
%
%
%-------------------------------------------------------------------------------------------------------------------------------------------
%

  \section{Comet 96P/Machholz 1 and the ETNOs}   
\begin{figure}
       \centering
        \includegraphics[width=\linewidth]{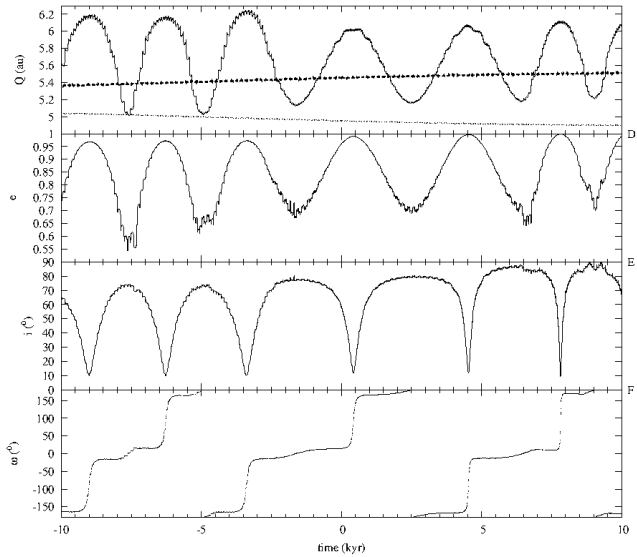}
        \caption{Aphelion distance, $Q$, panels D, E and F (of Fig. \ref{comet}) restricted to the time interval (-10, 10) kyr. For this 
                 object, when the eccentricity is at its highest, the inclination is at its lowest, and the argument of perihelion is close 
                 to 0\degr. In addition, its $Q$ librates around the value of Jupiter's own aphelion, thick curve, not perihelion, thin 
                 curve. 
                }
        \label{zoom}
     \end{figure}

\begin{figure}
       \centering
        \includegraphics[width=\linewidth]{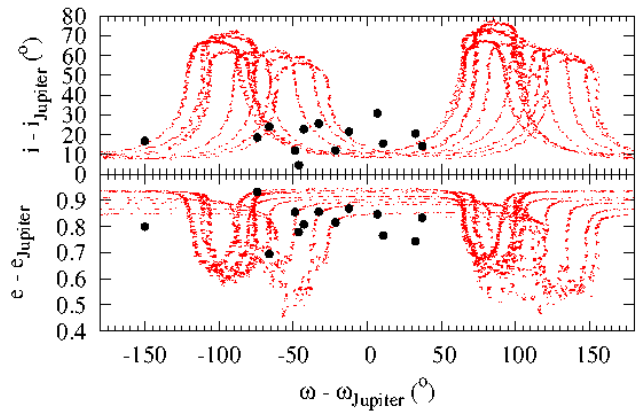}
        \caption{The $i_{\rm r}/e_{\rm r} \omega_{\rm r}$-portrait for comet 96P/Machholz 1 relative to Jupiter during the time interval
                 (-50, 0) kyr for the nominal orbit in Figs \ref{comet} and \ref{zoom}. The $i/e, \omega$-values for the ETNOs (see Table 
                 \ref{ETNOs}) are also plotted.
                }
        \label{key}
     \end{figure}

     The orbital evolution of comet 96P/Machholz 1 exhibits another unusual feature, when the eccentricity is at its highest, the 
     inclination is at its lowest, and the argument of perihelion is close to 0\degr (see Fig. \ref{zoom}). This interesting property may be 
     the key to explain the apparent lack of objects with $\omega$ around 180\degr found among known members of the ETNO population (see 
     above). ETNOs can only be discovered at perihelion and if they follow an orbital evolution similar to that of comet 96P/Machholz 1, 
     they will be preferentially found when their perihelion distance is the shortest; therefore, their eccentricity is the highest possible 
     which in turn implies that their $\omega$ must be close to 0\degr at the time of discovery. Perhaps, the secular evolution of these 
     objects is better viewed in the $e_{\rm r} \omega_{\rm r}$-plane, where $e_{\rm r} = e - e_{\rm p}$ and $\omega_{\rm r} = \omega - 
     \omega_{\rm p}$, $e_{\rm p}$ and $\omega_{\rm p}$ are, respectively, the eccentricity and argument of perihelion of a given planet 
     (Namouni 1999). In Fig. \ref{key}, we plot the $i_{\rm r}/e_{\rm r} \omega_{\rm r}$-map for comet 96P/Machholz 1 relative to Jupiter 
     when its dynamical state is the one of interest here. The $\omega, i/e$-values for the ETNO population (see Table \ref{ETNOs}) are also 
     plotted. The distribution of the orbital parameters ($\omega, i/e$) of most, if not all, ETNOs matches well what is expected of a Kozai 
     librator like comet 96P/Machholz 1. 
                 
     The Kozai scenario described for comet 96P/Machholz 1 is driven by the fact that the value of the aphelion distance of this comet 
     oscillates around that of Jupiter (see  Fig. \ref{zoom}, top panel), the perturber that controls its dynamical state. In other words, 
     for this mechanism to work, the aphelion of a massive perturber must be located relatively close to the aphelion distance of the 
     affected object and this putative perturber should probably move in a low-eccentricity, low-inclination orbit (see Fig. \ref{key}). If 
     we admit that comet 96P/Machholz 1 is a good dynamical analogue for all (or part) of the ETNO population, the implications for the 
     actual architecture of the trans-Plutonian region are multiple. First of all, massive perturbers must be orbiting near the aphelion 
     distances of the ETNOs in order to keep them locked in their current state. In addition, some (if not all) of these objects could be 
     transient. In Table \ref{ETNOs}, we observe four groups of aphelia. Asteroids 2003~HB$_{57}$, 2005~RH$_{52}$ and 2010~VZ$_{98}$ have 
     aphelia in the range 264--287 au. Asteroids (82158) 2001 FP$_{185}$, (148209) 2000~CR$_{105}$, 2002~GB$_{32}$, 2003~SS$_{422}$, 
     2007~VJ$_{305}$ and 2012~VP$_{113}$ have aphelia in the range 346--416 au. Asteroids 2004~VN$_{112}$ and 2010~GB$_{174}$ have aphelia 
     in the range 610--692 au. Finally, (90377) Sedna (2003~VB$_{12}$) and 2007~TG$_{422}$ have aphelia in the range 972--1001 au. If these 
     objects are experiencing a Kozai state analogous to that of comet 96P/Machholz 1, massive perturbers must be located at those 
     distances. The dispersion in aphelia can be used to constrain the semimajor axis/eccentricity of the perturber. Our analysis cannot 
     produce robust boundaries for the masses of these hypothetical perturbers but it is certain that at least several Earth masses are 
     required to maintain such Kozai states.  
%
%---------------------------------------------------------------------------------------------------------------------------------- FIGURE 1
%
%
%-------------------------------------------------------------------------------------------------------------------------------------------
%

     This comet may be just a limited dynamical analogue to 2012 VP$_{113}$ and similar objects because it mostly moves within the inner 
     Solar system; in contrast, the paths of ETNOs never get closer than that of Pluto. If most of the known ETNOs are experiencing Kozai 
     episodes of the type that characterises the evolution of this comet, the associated orbital signature would resemble what is currently 
     observed in the trans-Plutonian region. In addition, such scenario may explain naturally how objects originally located beyond 100 au 
     can eventually reach the trans-Neptunian region or even the realm of the giant planets (i.e. Centaurs) when their eccentricities become 
     too high. Under such a mechanism, retrograde Centaurs (see e.g. de la Fuente Marcos \& de la Fuente Marcos 2014a) may also be produced.
%
%-------------------------------------------------------------------------------------------------------------------------------------------
%
%
%-------------------------------------------------------------------------------------------------------------------------------------------
%

  \section{Conclusions}
     In this paper, we have studied the short-term dynamical evolution of orbital solutions similar to that of present-day comet 
     96P/Machholz 1 neglecting the effects of non-gravitational forces. This study has been carried out under the Newtonian and 
     post-Newtonian approximations. In both cases, we have found that the evolution of these orbits is unlike that of any other members of 
     the NEO population. Two rare evolutionary traits are observed: production of retrograde orbits and Kozai oscillations with zero 
     argument of perihelion at minimum perihelion distance. The observed pathway towards the retrograde orbital domain joins the mechanism
     presented by Greenstreet et al. (2012) in providing suitable scenarios to produce retrograde NEOs and rocky, high-velocity meteoroids
     on equally retrograde orbits. Furthermore, comet 96P/Machholz 1 appears to be a good dynamical match to most known members of the ETNO 
     population. It exhibits multiple properties that explain naturally the orbital distributions of these trans-Plutonian objects. The 
     incarnation of the Kozai mechanism driving the evolution of comet 96P/Machholz 1 and studied here opens a new and unexpected window 
     into the trans-Plutonian region that, if confirmed by the discovery of new members of the ETNO population exhibiting similar orbital 
     features, can change our current views of the outermost Solar system. However, the hypothesis of the existence of one or more massive 
     planets orbiting the Sun in nearly circular orbits so distant from Neptune is difficult to reconcile with the currently accepted 
     cosmogonic paradigm. It remains to be tested if a Kozai scenario such as the one described here may work when the massive perturbers 
     move in highly eccentric orbits, characteristic of scattered bodies (e.g. Bromley \& Kenyon 2014). 

  \section*{Acknowledgements}
     The authors thank the referee, J. A. Fern\'andez, for his prompt, constructive and helpful report. This work was partially supported by 
     the Spanish `Comunidad de Madrid' under grant CAM S2009/ESP-1496. CdlFM and RdlFM thank M. J. Fern\'andez-Figueroa, M. Rego Fern\'andez 
     and the Department of Astrophysics of the Universidad Complutense de Madrid (UCM) for providing computing facilities. Part of the 
     calculations and data analysis were completed on the `Servidor Central de C\'alculo' of the UCM. In preparation of this paper, we made 
     use of the NASA Astrophysics Data System, the ASTRO-PH e-print server and the MPC data server.

\end{document}